\documentclass[reprint,amsmath,amssymb,aps,superscriptaddress]{revtex4-1}

\usepackage{graphicx}
\usepackage{dcolumn}
\usepackage{bm}
\usepackage{color}

\begin{document}

\title{Magnetic properties of double perovskite $Ln_2$CoIrO$_6$ ($Ln$ = Eu, Tb, Ho): hetero-tri-spin $3d$-$5d$-$4f$ systems}

\author{Xiaxin Ding}
\affiliation{National High Magnetic Field Laboratory (NHMFL), Materials Physics and Applications (MPA)-Magnet (MAG) Group, Los Alamos National Laboratory (LANL), Los Alamos, New Mexico 87545, USA}
\author{Bin Gao}
\affiliation{Rutgers Center for Emergent Materials and Department of Physics and Astronomy, Piscataway, New Jersey 08854, USA}
\author{Elizabeth Krenkel}
\author{Charles Dawson}
\author{James C. Eckert}
\affiliation{Department of Physics, Harvey Mudd College, Claremont, California 91711, USA}
\author{Sang-Wook Cheong}
\affiliation{Rutgers Center for Emergent Materials and Department of Physics and Astronomy, Piscataway, New Jersey 08854, USA}
\author{Vivien Zapf}
\email{vzapf@lanl.gov}
\affiliation{National High Magnetic Field Laboratory (NHMFL), Materials Physics and Applications (MPA)-Magnet (MAG) Group, Los Alamos National Laboratory (LANL), Los Alamos, New Mexico 87545, USA}

\date{\today}

\begin{abstract}
The field of double perovskites is now advancing to three magnetic elements on the A, B and B$'$ sites. A series of iridium-based double perovskite compounds, $Ln_2$CoIrO$_6$ ($Ln$ = Eu, Tb, Ho) with three magnetic elements were synthesized as polycrystalline samples. The compounds crystalize in monoclinic structures with the space group $P2_1/n$. Magnetic properties of these hetero-tri-spin $3d$-$5d$-$4f$ systems were studied by magnetic susceptibility and field dependent magnetization in both DC and pulsed magnetic fields. All these compounds show ferrimagnetic transitions at temperatures $T_C$ above 100 K, which are attributable to antiferromagnetic coupling between Co$^{2+}$ and Ir$^{4+}$ spins. For Eu$_2$CoIrO$_6$, the magnetic properties are similar to those of La$_2$CoIrO$_6$. The Eu$^{3+}$ spins show Van Vleck paramagnetism that don't significantly interact with transition-metal cations. By contrast, Tb$_2$CoIrO$_6$ and Ho$_2$CoIrO$_6$ reveal a second transition to antiferromagnetic order below a lower temperature $T_N$. The temperature-induced ferrimagnetic-to-antiferromagnetic phase transition might be explained by a spin-reorientation transition. Moreover, a magnetic-field-induced spin-flop transition with a small hysteresis was observed below $T_N$ in these two compounds. The magnetic moment of all three compounds do not saturate up to 60 T at low temperatures. Moderate magnetocaloric effect was also observed in all three compounds. Our results should motivate further investigation of the spin configuration on single crystals of these iridium-based double perovskites.

\begin{description}
\item[PACS numbers]
75.50Ee, 75.50.Gg, 75.50.Lk
\end{description}
\end{abstract}

\pacs{Valid PACS appear here}

\maketitle


\section{INTRODUCTION}

The A$_2$BB$'$O$_6$ double perovskite family is a focus area of magnetic research due to a wide range of magnetic, magnetocaloric and multiferroic properties that reflect the design flexibility and interplay between charge, spin and lattice in these materials.~\cite{review}. The magnetic phases are controlled by the choice of magnetic or nonmagnetic cations on A, B and B$'$ sites in these compounds. In general, the A site is occupied by an alkaline or lanthanides ($Ln$) cation and B/B$'$ are transition metal elements. For a single magnetic B/B$'$-site compound, the superexchange coupling between two nearest cations through intermediate oxygen takes part in the magnetic order. In case of two magnetic B/B$'$-site cations, the magnetic properties of A$_2$BB$'$O$_6$ is usually dominated by the magnetic coupling between the local spin moments on B and B$'$ sites. In these compounds, the ferromagnetic (FM) ordering in the B and B$'$ cation sublattices can be explained by the indirect B-O-B$'$-O-B exchange interaction~\cite{Serrate}. Moreover, it is possible to have magnetic cations on the A site as well~\cite{LMIO, LCMO}, i.e. in Nd$_2$NiMnO$_6$, where the transition metal cations Ni$^{2+}$ and Mn$^{4+}$ order ferromagnetically to each other at 195 K, while the antiferromagnetic exchange between Nd$^{3+}$ and the transition metals arise at 50 K~\cite{NNMO}.

Another focus area for double perovskite research is Ir$^{4+}$ B or B$'$ cations. Ir$^{4+}$ provides strong and unusual spin-orbit-lattice coupling due to the comparable energy scales between spin-orbit coupling (SOC), on-site Coulomb interaction and crystal field energies~\cite{coupling}. For perfect octahedral symmetry, it is know that the 5$d$ levels of Ir should split into a $t_{2g}$ triplet and an $e_{g}$ doublet by the crystal electric field. Then, the strong SOC lifts the $t_{2g}$ orbital degeneracy to an effective $J_{eff} =1/2$ doublet ($e'$ level) and an effective $J_{eff} =3/2$ quartet ($u''$ level)~\cite{SOCL, SOCB}. For Ir$^{4+}$ (5$d^5$), the $t_{2g}$ level splits into a fully occupied $u''$ level and a half-filled $e'$ level, resulting in a total $J_{eff} = 1/2$ state. In La$_2$CoIrO$_6$ (monoclinic structure; $P2_1/n$), X-ray magnetic circular dichroism (XMCD) experiments confirm the valence states of the magnetic cations are Co$^{2+}$ and Ir$^{4+}$~\cite{XMCD12, XMCD15}, while distortions from perfect octahedral symmetry of the Ir$^{4+}$ B$'$ site can create a deviation from the  $J_{eff} = 1/2$ state. The temperature dependence of magnetization reveals magnetic order below $T_C$ = 95 K and the hysteresis loops at low temperatures indicate the presence of ferromagnetic (FM)-like components~\cite{neutron, XMCD12, XMCD15, Laspinglass}. Further studies have verified a ferrimagnetic (FiM) ground state in which a weak FM moment of canted Co$^{2+}$ spins is antiferromagnetically coupled to Ir$^{4+}$ cations with a negative moment~\cite{neutron, XMCD12, XMCD15}. This is explained in terms of the orbital hybridization of the high-spin (HS) Co$^{2+}$ $t_{2g}$ state and the Ir$^{4+}$ $J_{eff} = 1/2$ state. Recently, an interesting reentrant spin-glass magnetic behavior was observed in this compound~\cite{Laspinglass}. Naturally, magnetic A-site substitution modifies the magnetic properties as well as the structure of $Ln_2$CoIrO$_6$ double perovskites. Increased complexity of the magnetism is expected with interactions of three magnetic cations.

In this paper, we focus on the $Ln_2$CoIrO$_6$ ($Ln$ = Eu, Tb, Ho) family of double perovskites which is a hetero-tri-spin $3d$-$5d$-$4f$ system. Polycrystalline samples were synthesized. The symmetry of those double perovskites is compatible with the $P2_{1}/n$ space group. We performed a systematic investigation of these compounds through DC and pulsed magnetic susceptibility for the first time. All these compounds show a FiM transition at a high temperature $T_C$, which are attributable to antiferromagnetic (AFM) coupling between Co$^{2+}$ and Ir$^{4+}$ spins. In Eu$_2$CoIrO$_6$, the Van Vleck paramagnetic (PM) Eu$^{3+}$ cations don't interact with the transition-metal cations. Meanwhile the magnetic behaviors of Tb$_2$CoIrO$_6$ and Ho$_2$CoIrO$_6$ show a temperature-induced FiM-to-AFM phase transition and a field-induced spin-flop transition below $T_N$. A field up to 60 T is not enough to saturate their magnetic moments at low temperatures. Moderate magnetocaloric effect are observed around magnetic transitions for all three compounds.

\section{EXPERIMENTAL}

Polycrystalline samples of a series of lanthanide cobalt iridium oxides Eu$_2$CoIrO$_6$, Tb$_2$CoIrO$_6$ and Ho$_2$CoIrO$_6$ were synthesized using the conventional solid-state-reaction method. Stoichiometric Eu$_2$O$_3$ (99.9 \%), Tb$_4$O$_7$ (99.9 \%), Ho$_2$O$_3$ (99.9 \%), CoO (99.9 \%) and Ir (99.9 \%) powders were mixed, ground, pelleted and sintered at 1100 $^{\circ}$C, 1200 $^{\circ}$C and 1270 $^{\circ}$C for the first, second and third sintering, respectively. Powder X-ray diffraction (XRD) measurements were collected on a Rikagu X-ray diffraction instrument. The XRD data were analyzed by Rietveld refinement using the software MAUD~\cite{maud}. Magnetization was carried out between 2 and 300 K using a 14 T Quantum Design Physical Property Measurement System (PPMS) with a vibrating sample magnetometry (VSM) option in applied DC magnetic fields and in a 7 T Magnetic Property Measurement System (MPMS) Superconducting Quantum Interference Device (SQUID). The pulsed field magnetization measurements up to 60 T were performed at NHMFL in Los Alamos, in which the temperature down to 1.36 K was controlled with a $^4$He system. The pulsed field magnetization data were calibrated against DC measurements. 

\begin{figure}
\includegraphics[width=8.5 cm]{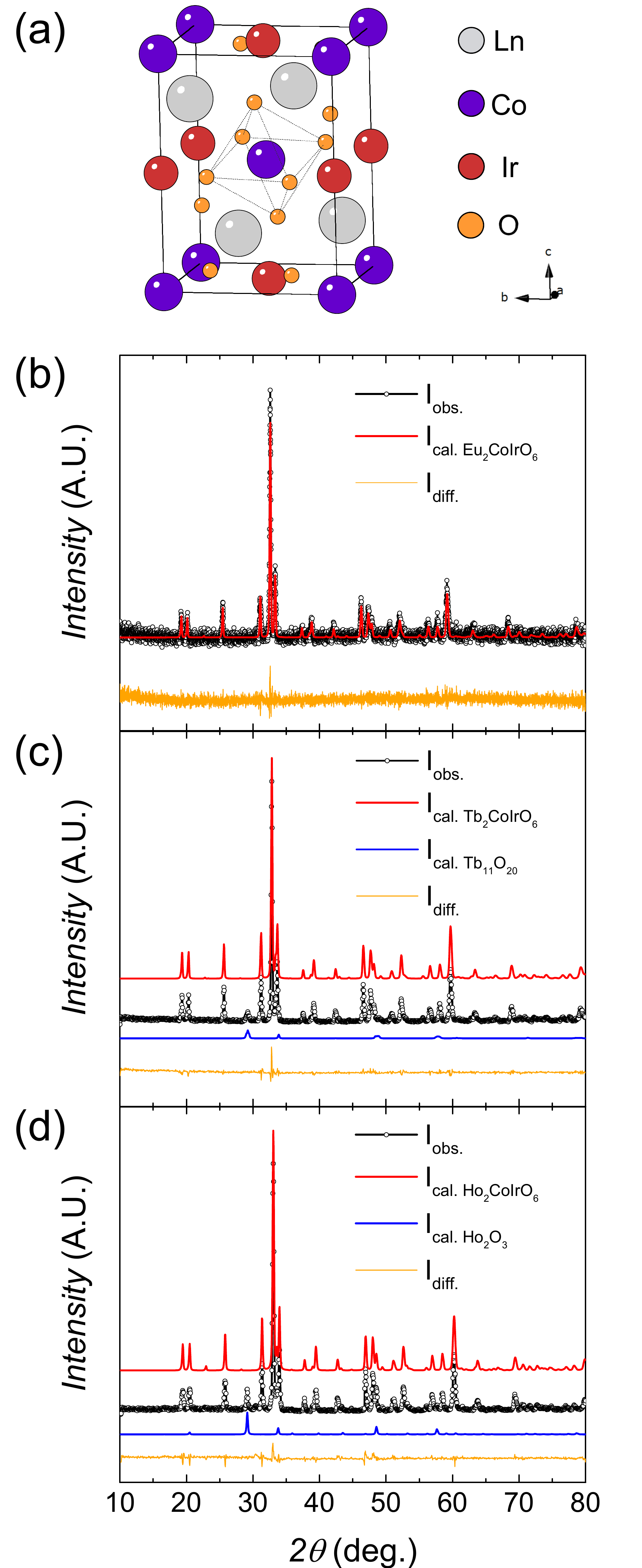}
\caption{\label{fig:fig1}(color online) (a) The structure of double perovskite $Ln_2$CoIrO$_6$. Powder XRD patterns of (b) Eu$_2$CoIrO$_6$, (c) Tb$_2$CoIrO$_6$ and (d) Ho$_2$CoIrO$_6$, shown in black. Red curves show the calculated pattern of the main phase $Ln_2$CoIrO$_6$ with the space group $P2_1/n$. Extra minor peaks are from the impurity phase, the calculated pattern shown in blue.} 
\end{figure}

\section{RESULTS AND WORKING MODEL}

\subsection{Structure}
Figure 1(a) displays the general structure of double perovskite $Ln_2$CoIrO$_6$. The corner-shared CoO$_6$ and IrO$_6$ octahedra alternate along three directions of the crystal, which form two monoclinic sublattices. The $Ln$ cations occupy the voids between the octahedra. The powder XRD patterns of Eu$_2$CoIrO$_6$, Tb$_2$CoIrO$_6$ and Ho$_2$CoIrO$_6$ measured at room temperature are shown in Fig. 1(b-d), which look very similar to that of Eu$_2$NiIrO$_6$ with the monoclinic structure~\cite{LMIO}. For Eu$_2$CoIrO$_6$, Rietveld refinement shows that this compound is a single-phase system with the space group $P2_{1}/n$, based on the Eu$_2$MgIrO$_6$ structure data~\cite{EMIO}. For Tb$_2$CoIrO$_6$, in addition to the Tb$_2$CoIrO$_6$ phase, a minor impurity phase of Tb$_{11}$O$_{20}$ (space group $P\bar{1}$~\cite{Tb11O20}) with a volume fraction of 6\% was found. Tb$_{11}$O$_{20}$ is AFM with a N\'eel temperature of 5.1 K~\cite{Tb11O20m}. For Ho$_2$CoIrO$_6$, less than 8.6\% of Ho$_2$O$_3$ (space group $Ia$-3~\cite{Ho2O3}) was present in the sample. Ho$_2$O$_3$ shows a second-order AFM transition with a N\'eel temperature of 2 K~\cite{Ho2O3m}. Structural parameters of these samples are listed in Table I. The cation radius decreases in the following order: La$^{3+}>$Eu$^{3+}>$Tb$^{3+}>$Ho$^{3+}$. The Rietveld fitting results reveal that the monoclinic structure is more distorted as the size of the $Ln$ cation becomes smaller. As can be expected, the structural distortion will result in a change of Co-O-Ir bond angles, which correlate with the magnetic ordering temperature.

\begin{figure}
\includegraphics[width=7.6 cm]{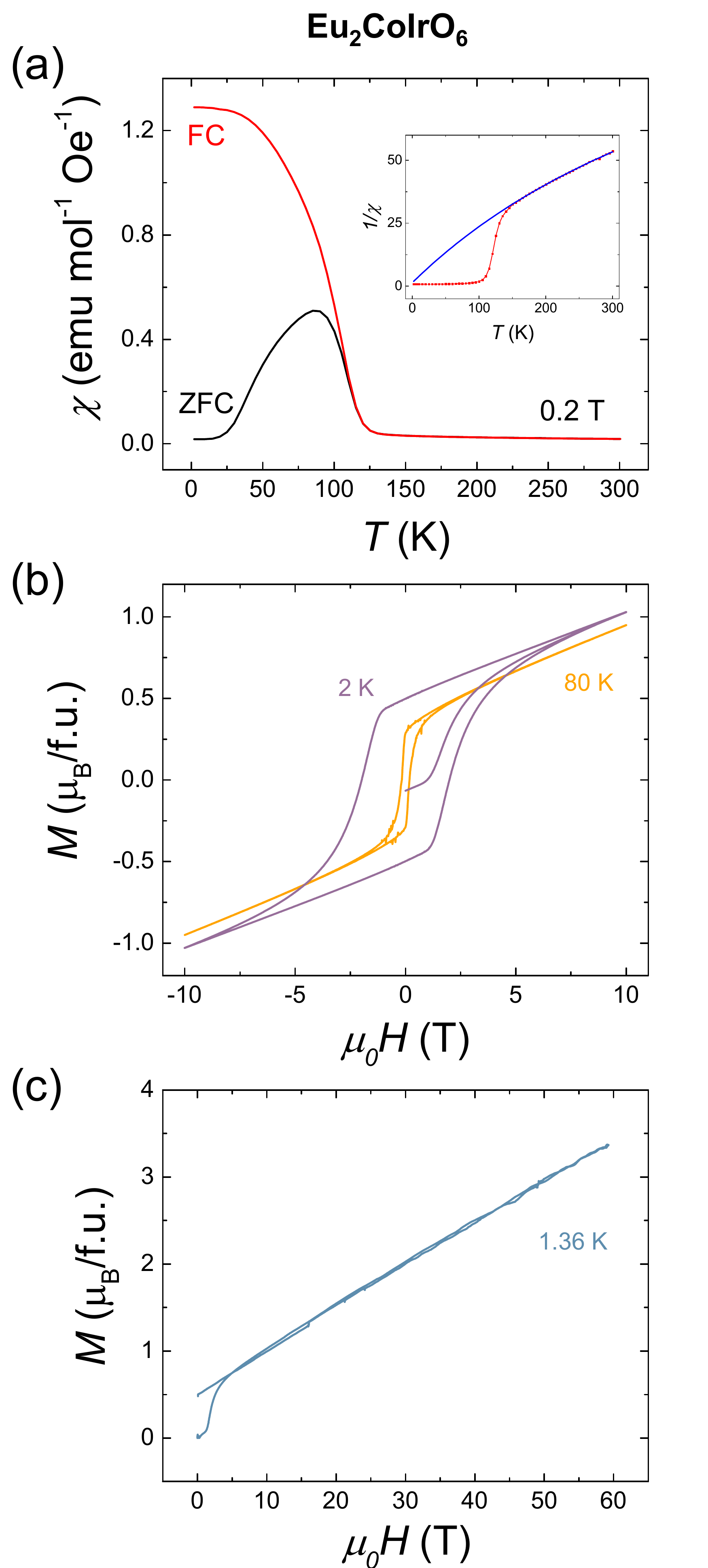}
\caption{\label{fig:fig2}(color online) Magnetic properties of Eu$_2$CoIrO$_6$. (a) Temperature dependence of ZFC and FC magnetic susceptibility $\chi(T) = M/H$ at $H$ = 0.2 T. Inset: inverse FC data $1/\chi(T)$. The blue solid line is the modified Curie-Weiss fit. (b) Isothermal curves of magnetization verses DC magnetic field at various temperatures. (c) Pulsed field magnetization data up to 60 T at 1.36 K.}
\end{figure}

\subsection{Magnetism}

\subsubsection{Eu$_2$CoIrO$_6$} 

Temperature dependent zero-field-cooled (ZFC) and field-cooled (FC) DC magnetic susceptibility $\chi(T)$ measurements of Eu$_2$CoIrO$_6$ performed at 0.2 T are presented in Fig. 2(a). The derivative of the FC susceptibility $d\chi/dT$ reveals a FM-like magnetic transition below the Curie temperature $T_C$ = 105 K. Compared to La$_2$CoIrO$_6$, the Eu substitution of La at the A site results in a substantial shift of the $T_C$ to a higher temperature. The large contrast between the ZFC and FC data indicates the presence of FM-like components, which is confirmed by the hysteresis in Fig. 2(b). A plateau is clearly observed below $T_C$ in the FC curve. A peak exists at $T_p$ = 85 K in the ZFC curve. $T_p$ shifts to lower temperatures with increasing the external magnetic field, which is not shown here. 1/$\chi$ at high temperatures violates the linear Curie-Weiss law. Instead, the magnetic susceptibility follows a modified Curie-Weiss law with $\chi = \chi_0+C/(T-\theta)$ above 160 K, where $\chi_0$ is a fitting term and arises mainly from Eu$^{3+}$, $C$ is the Curie constant and $\theta$ is the  Curie-Weiss temperature. This form of susceptibility is consistent with Van Vleck paramagnetism~\cite{Vleck}, which is often seen in europium containing compounds~\cite{EuBO3, EuOOH, LMIO}. As shown in the inset of Fig. 2(a), the fit gives a Curie-Weiss temperature of $\theta$ = -5.3 K, however we note that in these materials single-ion anisotropies and level splittings can influence or even dominate the Curie-Weiss temperature, in addition to AFM and FM interactions. The effective moment $\mu_{eff}$ = 5.5 $\mu_B$/f.u. is calculated with $\chi_0$ taken out. As shown in Fig. 2(b), the remanent magnetization of Eu$_2$CoIrO$_6$ at 2 K is found to be 0.5 $\mu_B$ per formula unit (f.u.) which is smaller than 0.7 $\mu_B$/f.u. of La$_2$CoIrO$_6$ at 5 K~\cite{Laspinglass}. The step-like magnetic transition happens at $H_c$ = 1.3 T in the initial hysteresis loop, which has been observed in La$_2$CoIrO$_6$~\cite{Laspinglass}.

As its behavior is similar to that of the reference compound La$_2$CoIrO$_6$~\cite{XMCD15, Laspinglass}, we can hypothesis that the PM Eu$^{3+}$ doesn't interact with the other two magnetic cations, and the ground state of the Co$^{2+}$ and Ir$^{4+}$ is FiM~\cite{XMCD12, XMCD15}. The peak in the ZFC curve below $T_C$ and the step-like magnetic transition in the $M(H)$ curve are characteristics of AFM-FM evolutions seen in similar materials such as Lu$_2$CoMnO$_6$~\cite{LMCO}, and have also been attributed to spin-glass-like states in La$_2$CoIrO$_6$~\cite{Laspinglass}. The strong linear contribution beyond the hysteresis loop is most likely due to the gradual field alignment of the canted Co$^{2+}$ and Ir$^{4+}$ magnetic moments away from the easy axis~\cite{neutron}. The effective moment of the HS Co$^{2+}$ has been reported to be 4.8 $\mu_B$ in related double perovskites, which is higher than the spin-only value of 3.87$\mu_B$ due to the partial unquenched orbital contribution~\cite{YCMO}. The effective moment of Ir$^{4+}$ was reported to be 1.3 $\mu_B$ in La$_2$MgIrO$_6$~\cite{LMIO}. Therefore, the AFM coupling of Co$^{2+}$ and Ir$^{4+}$ should result in a saturated spin moment of  3.5 $\mu_B$/f.u., neglecting the paramagnetism of Eu$^{3+}$. However, as shown in Fig. 2(c), $M$ does not saturate even up to 60 T. $M(H)$ varies almost linearly with increasing field and reaches 3.37 $\mu_B$/f.u. at 60 T. Since the Co$^{2+}$ spin is not fully aligned at 60 T, the effective moment of Ir$^{4+}$ should be less than 1.43 $\mu_B$. Nevertheless, Kolchinskaya $et$ $al$ reported an unusually large total magnetic moment 0.38 $\mu_B$/f.u. for Ir$^{4+}$ in La$_2$CoIrO$_6$ by XMCD~\cite{XMCD12}. In this case, the saturation moment of Co$^{2+}$ and Ir$^{4+}$ should be 4.42 $\mu_B$/f.u., which needs to be confirmed by measurements in higher magnetic fields. Thus, the effective moment of Ir$^{4+}$ should be in the range of 0.4-1.4 $\mu_B$/f.u. in these double perovskites.

\begin{figure*}
\includegraphics[width=18 cm]{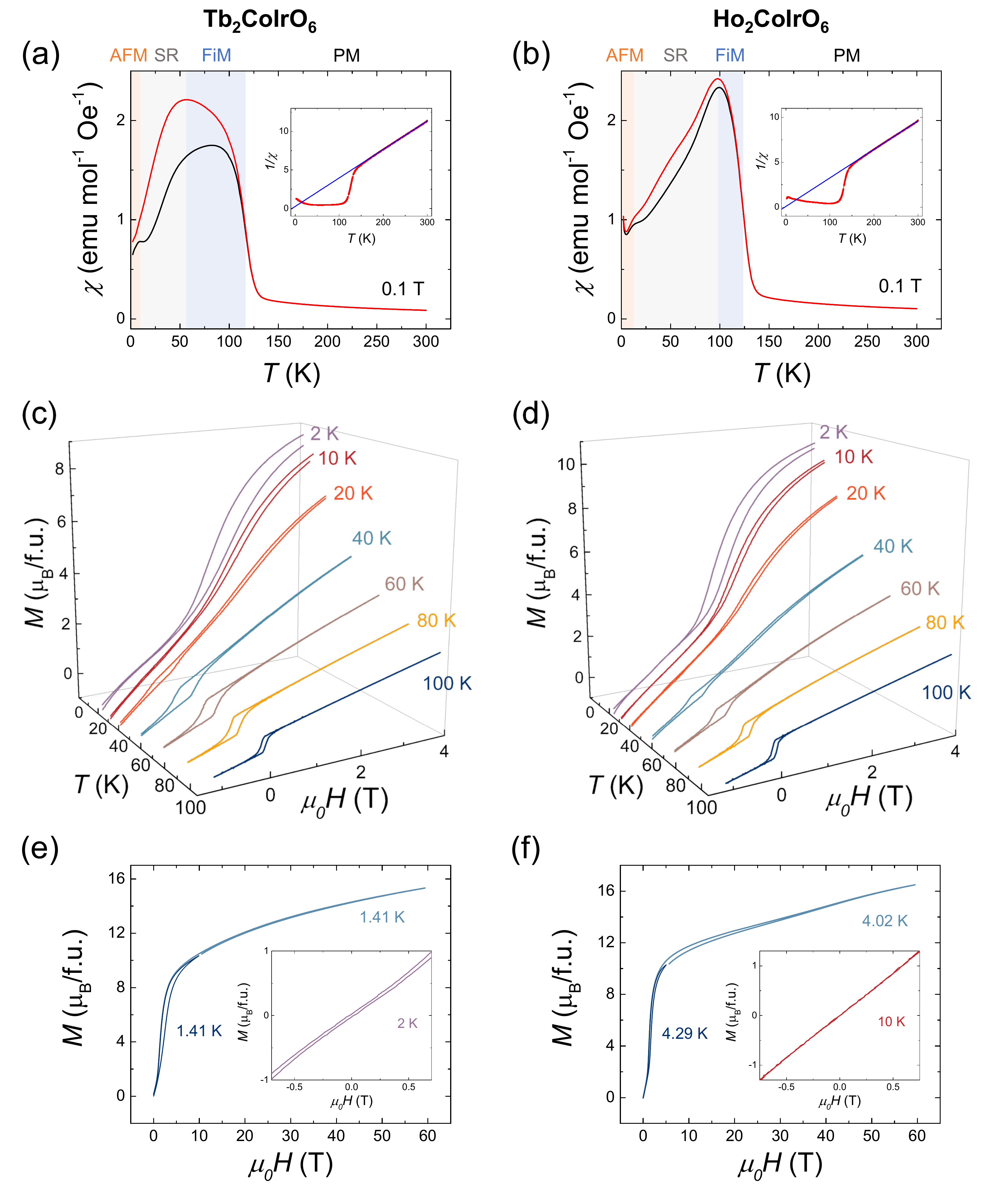}
\caption{\label{fig:fig3}(color online) Magnetic properties of Tb$_2$CoIrO$_6$ and Ho$_2$CoIrO$_6$. (a,b) Temperature dependence of the magnetic susceptibility  at 0.1 T. The change of 1/$\chi$ is shown in the inset. The blue solid line is the linear Curie-Weiss fit. Colored background in the main frame shows the temperature region of FiM, SR and AFM, while the uncolored is PM. (c,d) Magnetization verses DC magnetic field ranging from -1 to 4 T at various temperatures. The $M(H)$ data is plotted in 3D graphs to make a clear view. (e,f) Main frame: pulsed field magnetization data up to 60 T. Inset: amplified view of the $M(H)$ data below $T_N$ measured in DC fields.}
\end{figure*}

\subsubsection{Tb$_2$CoIrO$_6$} 

Fig. 3(a) displays the temperature dependence of susceptibility of Tb$_2$CoIrO$_6$. 1/$\chi$ above 185 K is well fitted by the Curie-Weiss expression, $\chi = C/(T-\theta)$. We report a Curie-Weiss temperature of -7.1 K, however as noted previously this temperature can be strongly influenced by factors besides magnetic exchange interactions. The effective magnetic moment of Tb$_2$CoIrO$_6$ obtained by the fitting is $\mu_{eff}^{exp} = \sqrt{8C}$ = 14.70 $\mu_B$/f.u., where $k_B$ is the Boltzmann constant. The expected moment of Tb$^{3+}$ is 9.72 $\mu_B$, calculated by $\mu_{Tb} = g_{Tb}\sqrt{J_{Tb}(J_{Tb}+1)}$ where $g_{Tb}$ is the Land\'e $g$-factor. Based on the value of $\mu_{Ir}$ determined above, the expected effective moment of Tb$_2$CoIrO$_6$ can be calculated to be 14.57-14.63 $\mu_B$/f.u. according to $\mu_{eff} = \sqrt{2\mu_{Tb}^2+\mu_{Co}^2+\mu_{Ir}^2}$, which is close to but slightly less than the value obtained experimentally. A sudden jump at $T_{C}$ = 117 K signals the onset of a FM-like ordering. The hysteresis below 100 K in Fig. 3(c) could be explained by the FiM ordering due to the AFM coupling between the canted Co$^{2+}$ and Ir$^{4+}$ as was suggested for La$_2$CoIrO$_6$ and for Eu$_2$CoIrO$_6$. The ZFC and FC curves also separate below $T_C$ with a peak at 82 K in the ZFC curve, similar to Eu$_2$CoIrO$_6$ and La$_2$CoIrO$_6$. As shown in Fig. 3(c), it is clear that the remanent moment and the coercive field increase with decreasing temperature from 100 to 60 K. Beyond the FiM hysteresis, the magnetization increases nonlinearly with increasing field, indicating a component of paramagnetism of Tb$^{3+}$.

As the temperature decreases further, a downturn occurs clearly at 56 K in the FC curve. Moreover, an AFM transition is signified by a kink in the ZFC curve around 10 K. Since the spins of magnetic rare earth cations usually order at low temperatures, the magnetic behavior below $T_N$ = 10 K might be strongly affected by the alignment of Tb$^{3+}$ spins. The AFM ordering is confirmed by $M(H)$ curves below 10 K in Fig. 3(c). At 2 K, $M$ increases almost linearly with the magnetic field at the beginning, then undergoes a substantial increase at $H_m$ = 2.4 T. Thus, a metamagnetic transition occurs. A hysteresis is observed in the metamagnetic transition, implying a weak first-order transition. It is worthwhile to note that there is a tiny remanent moment of ~0.02 $\mu_B$/f.u. at 2 K after the magnetic field is turned off, as shown in the inset of Fig. 3(e). This might come from the impurity phase Tb$_{11}$O$_{20}$ which shows a remanence in the $M(H)$ curve at 1.9 K~\cite{Tb11O20m}. Generally, in an AFM system, the metamagnetic transition corresponds to a spin-flop transition from an AFM state to a spin ferromagnetically polarized state~\cite{flop}. 

Especially interesting is that this compound exhibits a FiM to AFM phase transition with decreasing temperature, which might be caused by a change of the AFM structure type or by the spin-reorientation (SR) transition~\cite{reorientation}. Now let's focus on the temperature region from 10 to 56 K. Below 56 K, the coercive field of the FiM hysteresis continues to increase with decreasing temperature until it becomes undistinguishable at 10 K, while the remanence starts to decrease with decreasing temperature. The change of the FiM hysteresis indicates that (i) the AFM coupling between Co$^{2+}$ and Ir$^{4+}$ doesn't change, and (ii) The FM component of Co$^{2+}$ becomes smaller as temperature decreases. Meanwhile, the metamagnetic hysteresis emerges in this temperature region, which is clear at 20 K. Since the $M(H)$ of polycrystalline samples is the average of magnetic moments for different directions, one possible scenario is that the Co spins reoriented and antiferromagnetically ordered along the other direction. This temperature-induced orientational transition can be caused by competing anisotropy of transition metal sublattices. Because of the impurity phase, it's hard to tell whether the magnetic moments of Co$^{2+}$ and Ir$^{4+}$ are compensated below $T_N$. 

The pulsed field magnetization of Tb$_2$CoIrO$_6$ was carried out at 1.41 K. Since the signal voltage from the coil is proportion to $dM/dt$, the sharp transition at $H_m$ results a large voltage which saturated the data acquisition system in a 60 T shot. We solve this problem by combining 10 T data (no saturation problem at 2.4 T) and 60 T (saturation problem near 2.4 T) in Fig. 3(e). Beyond the hysteresis, $M$ increases nonlinearly and does not saturate up to 60 T. The moment reaches 15.33 $\mu_B$/f.u. at 60 T.

\subsubsection{Ho$_2$CoIrO$_6$} 

Ho$_2$CoIrO$_6$ shows similar magnetic properties as Tb$_2$CoIrO$_6$. It undergoes a FiM transition at 123 K followed by a divergence between the ZFC and FC curves shown in Fig. 3(b). The inverse of susceptibility presents a linear behavior at high temperature. A Curie-Weiss fit above 185 K gives a Curie-Weiss temperature of $\theta$ = -2.6 K. The obtained $\mu_{eff}^{exp} = 15.87 \mu_B$ from the fit is in reasonably good agreement with the theoretical moment 15.75-15.80 $\mu_B$. Broad peaks are observed at 99 K in ZFC curve and 98 K in FC curve which might indicate the SR transition. The kink at 13 K indicates the AFM transition. The increase of susceptibility below 5 K is presumably attributed to the existence of Ho$_2$O$_6$ impurity, which has a strong Curie-Weiss increase at low temperatures. Isothermal magnetization curves are shown in Fig. 3(d). Below $T_N$, the magnetization increases linearly in weak fields, as shown in the inset of Fig. 3(d). The field-induced spin-flop transition happens at $H_m$ with a small hysteresis. The inset of Fig. 3(f) shows the amplified view of the $M(H)$ curve at 10 K. It is clear that there is no remanence. Thus, the FiM moment vanishes below $T_N$. According to our working model, the SR transition happens in a region from 13 to 98 K. Moreover, the magnetic moments of Co$^{2+}$ and Ir$^{4+}$ are compensated below $T_N$. The main frame of Fig. 3(f) shows further pulsed field magnetization data. The transition at $H_m$ was so sharp that even a 10 T shot saturated the data acquisition system. Thus, the pulsed field data are a combined plot of 5 T and 60 T shots. There is no saturation trend up to 60 T at 4.02 K.

\begin{figure}
\includegraphics[width=8 cm]{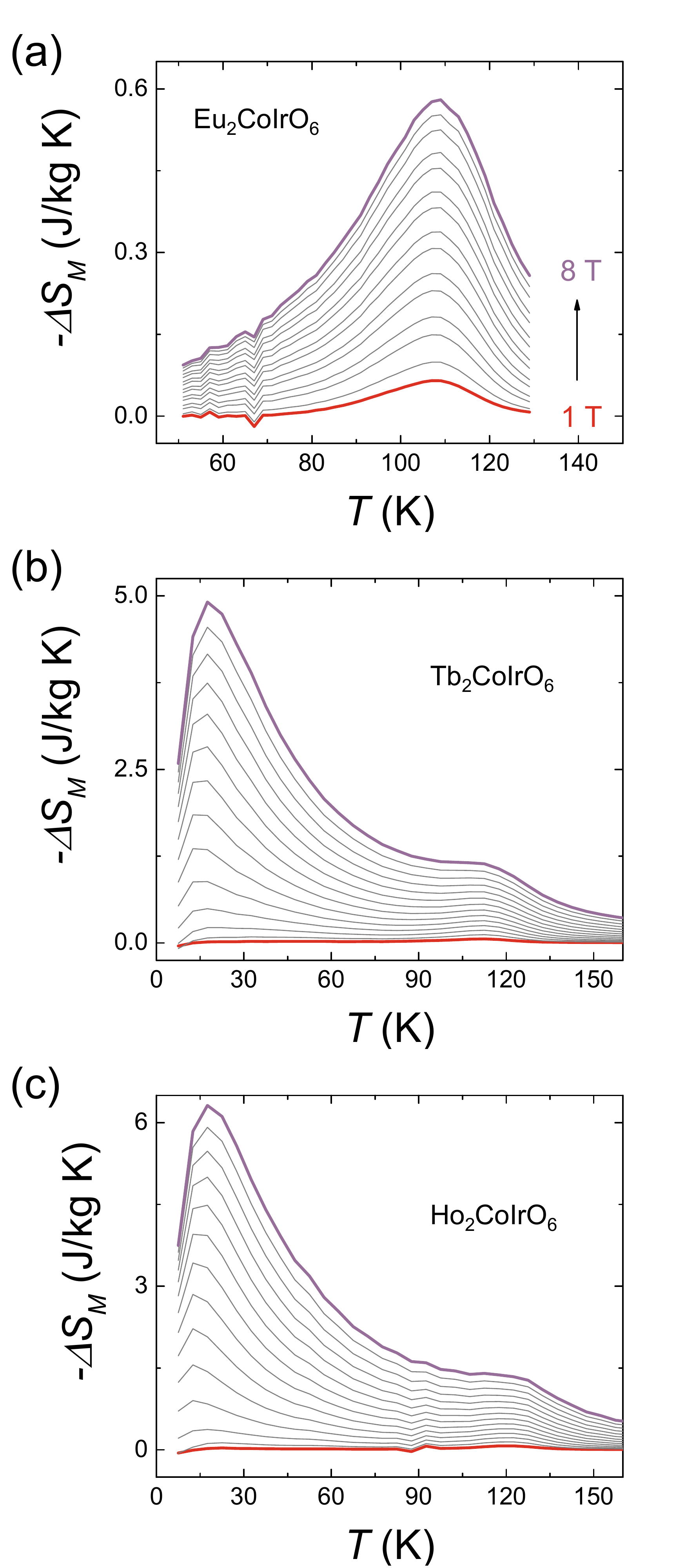}
\caption{\label{fig:fig4}(color online) Thermal profile of field induced magnetic entropy change $-\Delta S_M$ under the applied field changing form 1 to 8 T for (a) Eu$_2$CoIrO$_6$, (b) Tb$_2$CoIrO$_6$ and (c) Ho$_2$CoIrO$_6$, respectively.}
\end{figure}

\subsection{Magnetocaloric effect}

The magnetocaloric effect (MCE) is defined as the adiabatic temperature change $\Delta T$ or isothermal magnetic entropy change $\Delta S_M$ of a magnetic material due to a varying external magnetic field~\cite{CPB}. The magnetic entropy change $\Delta S_M$ can be measured directly with the calorimetry method or indirectly calculated from magnetization measurements using Maxwell's thermodynamic relation: 
\begin{eqnarray}
\nonumber \Delta S_M(T,H) = \int_{0}^{H_1}\left(\frac{\partial M(T,H)}{\partial T}\right)_H\mathrm{d}H.
\end{eqnarray}
$-\Delta S_M$ usually reaches a maximum around the magnetic transition temperature, such as $T_C$. In order to understand the field dependent magnetic behavior of the three compounds and determine their magnetocaloric potential, $M(H)$ curves of these samples were measured at various temperatures. Fig. 4 summarizes the temperature dependence of $-\Delta S_M$ of $Ln_2$CoIrO$_6$ samples obtained at different magnetic field changes (from 1 T to 8 T). They all show a moderate MCE and a peak around $T_C$. For Eu$_2$CoIrO$_6$, $-\Delta S_M$ reaches 0.58 J/Kg-K at 8 T. For Tb$_2$CoIrO$_6$ and Ho$_2$CoIrO$_6$, in addition to the peak around $T_C$, $-\Delta S_M$ becomes negative for small fields below $T_N$ which indicates the presence of an AFM component. This behavior is known as the inverse MCE. As the field increased above $H_m$, the sign of $-\Delta S_M$ changes to positive. The magnitude increases with increasing field and reaches a maximum around $T_N$. The peak value for Tb$_2$CoIrO$_6$ and Ho$_2$CoIrO$_6$ at 8 T are 4.91 and 6.32 J/Kg-K, respectively. The reversal of the sign of $-\Delta S_M$ at low temperatures is consistent with the spin-flop transition observed in $M(H)$ curves.

\begin{table*}
  \begin{center}
    \caption{Structural parameters, magnetic ordering temperatures, Curie-Weiss temperature and effective moment of double perovskites $Ln_2$B$^{2+}$Ir$^{4+}$O$_6$.}
    \label{tab:table1}
    \renewcommand{\arraystretch}{2}
    \begin{tabular}{m{2.6cm}m{1.5cm}m{1.5cm}m{1.5cm}m{1.5cm}m{1.5cm}m{1.2cm}m{1.2cm}m{1.2cm}m{2cm}}
    \hline \hline
      $Ln_2$BIrO$_6$ & $a$(\AA) & $b$(\AA) & $c$(\AA) & $\beta$(deg.) & $V$(\AA$^3$) & $T_C$(K) & $T_N$(K) & $\theta$(K) & $\mu_{eff}$($\mu_B$/f.u.)\\ 
      \hline
      La$_2$CoIrO$_6$~\cite{neutron} & 5.581(9)  & 5.657(6) & 7.907(8) & 89.98(1) & 249.7(3) & 95 & n/a &-13.9 & 4.71\\
      Eu$_2$CoIrO$_6$ & 5.365(9)  & 5.741(3) & 7.676(0) & 90.028(4) & 236.4(8) & 105 & n/a & -5.3 & 5.50\\ 
      Tb$_2$CoIrO$_6$ & 5.319(3) & 5.724(7) & 7.627(2) & 90.047(9) & 232.2(6) & 117 &10  & -7.1 & 14.70\\ 
      Ho$_2$CoIrO$_6$ & 5.271(6) & 5.697(1) & 7.577(4) & 90.14(80) & 227.5(7) & 123 & 13 & -2.6 & 15.87\\ 
      \hline
      La$_2$NiIrO$_6$~\cite{LNIO,LMIO} & 5.575(3)  & 5.626(1) & 7.898(1) & 90.02(7) & 247.7(4) & 85 & n/a & -28 & 3.28\\
      Nd$_2$NiIrO$_6$~\cite{LMIO}  & 5.433(0)  & 5.694(3) & 7.763(9) & 90.004(0) & 240.1(9) & 125 & 5 & -32 & 6.19\\
      Eu$_2$NiIrO$_6$~\cite{LMIO}  & 5.378(7)  & 5.715(0) & 7.706(1) & 90.035(0) & 236.8(8) & 162 & n/a & n/a & n/a\\
      Gd$_2$NiIrO$_6$~\cite{LMIO}  & 5.326(3)  & 5.718(9) & 7.651(5) & 90.026(0) & 233.0(7) & 170 & 8 & ? & 11.35\\
       \hline
      La$_2$MgIrO$_6$~\cite{LMIO,EMIO} & 5.599(7)  & 5.606(7) & 7.916(4) & 90.005(7) & 248.5(4) & n/a & 10 & -10 & 1.31\\
      Nd$_2$MgIrO$_6$~\cite{LMIO,EMIO} & 5.478(7)  & 5.651(7) & 7.812(5) & 90.021(1) & 241.9(1) & 9? & 5 & -19 & 4.84\\
      Eu$_2$MgIrO$_6$~\cite{EMIO} & 5.391(9)  & 5.679(7) & 7.725(8) & 90.059(1) & 236.(60) & n/a & 10 & n/a & n/a\\
      Gd$_2$MgIrO$_6$~\cite{EMIO} & 5.365(4)  & 5.687(1) & 7.701(4) & 90.193(3) & 234.9(9) & ? & n/a & 3 & 10.68\\
      \hline \hline
    \end{tabular}
  \end{center}
\end{table*}

\section{DISCUSSIONS}

Once we settle lanthanide cations on the A site of A$_2$BIrO$_6$, the oxidation states of B and Ir are allowed for combinations of +1/+5, +2/+4 and +3/+3. The combination of +3/+3 is quite rare. To the best of our knowledge, it was only found in La$_2$FeIrO$_6$~\cite{LFIO2}. In La$_2$FeIrO$_6$, the fully occupied $u''$ level and fully occupied $e'$ level of Ir$^{3+}$ (5$d^6$) lead to a nonmagnetic ground state ($J_{eff} = 0$), thus there are only superexchange AFM coupling between Fe$^{3+}$ cations~\cite{LFIO}. To achieve the combination of +1/+5, one way is to place alkali metal cations on the B site. As is the case of La$_2$LiIrO$_6$, the Ir has been found to be in the oxidation state of 5+~\cite{LLIO}. In case of Ir$^{5+}$ (5$d^4$), a nonmagnetic $J_{eff} = 0$ ground state is expected with a fully occupied $u''$ level. Therefore, no magnetic transition was found in La$_2$LiIrO$_6$. The combination of +2/+4 is a good playground for studying the magnetic coupling of 3$d$ transition metals to the 5$d$ Ir with strong spin-orbit coupling. In Table I, we listed the structural parameters and magnetic ordering temperatures of our three compounds and the reference compound La$_2$CoIrO$_6$, as well as the other $Ln_2$B$^{2+}$Ir$^{4+}$O$_6$ materials. It is clear that as the size of lanthanide cation becomes smaller, the lattice constants $a$ and $c$ decrease, enhancing the monoclinic distortion with decreasing the unit-cell volume monotonically. This behavior is consistent with the series of $Ln_2$NiIrO$_6$~\cite{LMIO} and $Ln_2$CoMnO$_6$~\cite{LCMO} compounds. The enhanced structural distortion will result in smaller B-O-B$'$ bond angles which correlate with the magnetic ordering temperature. In $Ln_2$CoMnO$_6$, the magnetic transition temperature decreases linearly with the enhancement of monoclinic distortion~\cite{LCMO}. However, in $Ln_2$NiIrO$_6$ and $Ln_2$CoIrO$_6$, $T_C$ increases gradually as decreasing the size of rare earth cations, as listed in Table I. Moreover, with magnetic $4f$ rare earth metals, there is an AFM transition at low temperatures in the hetero-tri-spin $3d$-$5d$-$4f$ system, which should be strongly influenced by rare earth magnetic moments. $T_N$ also increases with decreasing the size of magnetic rare earth cations. The temperature-induced FiM-to-AFM phase transition could be explained by the spin-reorientation transition scenario. Instead of placing 3$d$ transition metal cations on the B site, the other way to achieve the combination of +2/+4 for B/Ir is the choice of nonmagnetic alkaline earth metal cations on the B site. In La$_2$MgIrO$_6$~\cite{LMIO} and Eu$_2$MgIrO$_6$~\cite{EMIO}, the Ir$^{4+}$ cations order antiferromagnetically below 10 K. For Nd$_2$MgIrO$_6$ and Gd$_2$MgIrO$_6$, they belong to the hetero-bi-spin $5d$-$4f$ system which exhibits slightly more complex magnetic behavior due to the magnetic coupling between the $Ln^{3+}$ and Ir$^{4+}$ cations~\cite{EMIO}. There are two magnetic transitions in the susceptibility of Nd$_2$MgIrO$_6$ at 5 and 9 K. For Gd$_2$MgIrO$_6$, there is no AFM transition in the susceptibility.

The substitution of the trivalent rare earth cation by the divalent alkaline earth cation on the A site tends to change the Ir valence from 4+ to 5+. Since Ir$^{5+}$ cations are nominally nonmagnetic, no magnetic transition is found for Ir$^{5+}$ of the undistorted Sr$_2$YIrO$_6$ ($Fm3m$)~\cite{SYIO99, SYIO17}. By gradually substituting Sr on the A site of La$_2$CoIrO$_6$, the following phase transitions occur at room temperature: $P2_1/n \rightarrow P2_1/n+I2/m \rightarrow I2/m$~\cite{neutron}, as well as a change of valence state from Co$^{2+}$/Ir$^{4+}$ to Co$^{3+}$/Ir$^{5+}$~\cite{XMCD12}. Furthermore, the fully occupied $u''$ level of Ir$^{5+}$ is expected to hamper its magnetic coupling to the 3$d$ transition metal elements on the B site. This is exemplified in Sr$_2$CoIrO$_6$ ($I2/m$), Ir$^{5+}$ has a paramagnetic moment with almost no orbital contribution, meanwhile Co$^{3+}$ cations order antiferromagnetically without canting~\cite{XMCD12}. 

Finally, by controlling the oxygen deficiency of Sr$_2$CoIrO$_{6-\delta}$, a rhombohedral phase with the composition Sr$_3$CoIrO$_6$ can be formed~\cite{S3CIO}. It contains parallel one-dimensional chains along the $c$-axis~\cite{chains}. 

\section{CONCLUSIONS}

In this work, we reported the synthesis, crystal structures and magnetic behavior of iridium-based double perovskite $Ln_2$CoIrO$_6$ ($Ln$ = Eu, Tb, Ho) polycrystalline samples. These compounds crystalize in the monoclinic space group $P2_1/n$. All of the compounds exhibited FiM Co$^{2+}$-Ir$^{4+}$ interactions at high temperatures. The Eu$^{3+}$ spins show Van Vleck paramagnetism in Eu$_2$CoIrO$_6$ that don't show observed interaction with Co$^{2+}$ or Ir$^{4+}$. However, with magnetic Tb$^{3+}$ and Ho$^{3+}$ cations on the A site, a second AFM transition was observed at low temperatures. A magnetic-field-induced spin-flop transition with a small hysteresis occurred below $T_N$ in these two compounds. We used a spin-reorientaion working model to explain the temperature-induced FiM-to-AFM phase transition in this hetero-tri-spin $3d$-$5d$-$4f$ system. A field up to 60 T is not enough to saturate their magnetic moments at low temperatures. Finally, moderate magnetocaloric effect was observed around magnetic transitions for all three compounds. The intriguing magnetic properties of these compounds call for high-quality single crystals. Moreover, to further explore the spin configuration and verify our working model, the spin structure obtained from inelastic neutron scattering experiment will be needed.

\section{ACKNOWLEDGEMENTS}

This work is supported by the Laboratory-Directed Research and Development program at Los Alamos National Lab under the auspices of the U.S. Department of Energy (DOE). The National High Magnetic Field Lab Pulsed Field facility is supported by the National Science foundation under cooperative Grant Nos. DMR-1157490 and DMR-1644779, the U.S. DOE, and the State of Florida. The work at Rutgers University was supported by the NSF under Grant No. DMR-1629059. The work at Harvey Mudd College was supported by a grant from the Jean Perkins Foundation. We thank John Singleton for the use of his magnetization probe in pulsed field experiments.

\nocite{*}

\bibliography{DoublePerovskite}

\providecommand{\noopsort}[1]{}\providecommand{\singleletter}[1]{#1}%
\begin{thebibliography}{34}%
\makeatletter
\providecommand \@ifxundefined [1]{%
 \@ifx{#1\undefined}
}%
\providecommand \@ifnum [1]{%
 \ifnum #1\expandafter \@firstoftwo
 \else \expandafter \@secondoftwo
 \fi
}%
\providecommand \@ifx [1]{%
 \ifx #1\expandafter \@firstoftwo
 \else \expandafter \@secondoftwo
 \fi
}%
\providecommand \natexlab [1]{#1}%
\providecommand \enquote  [1]{``#1''}%
\providecommand \bibnamefont  [1]{#1}%
\providecommand \bibfnamefont [1]{#1}%
\providecommand \citenamefont [1]{#1}%
\providecommand \href@noop [0]{\@secondoftwo}%
\providecommand \href [0]{\begingroup \@sanitize@url \@href}%
\providecommand \@href[1]{\@@startlink{#1}\@@href}%
\providecommand \@@href[1]{\endgroup#1\@@endlink}%
\providecommand \@sanitize@url [0]{\catcode `\\12\catcode `\$12\catcode
  `\&12\catcode `\#12\catcode `\^12\catcode `\_12\catcode `\%12\relax}%
\providecommand \@@startlink[1]{}%
\providecommand \@@endlink[0]{}%
\providecommand \url  [0]{\begingroup\@sanitize@url \@url }%
\providecommand \@url [1]{\endgroup\@href {#1}{\urlprefix }}%
\providecommand \urlprefix  [0]{URL }%
\providecommand \Eprint [0]{\href }%
\providecommand \doibase [0]{http://dx.doi.org/}%
\providecommand \selectlanguage [0]{\@gobble}%
\providecommand \bibinfo  [0]{\@secondoftwo}%
\providecommand \bibfield  [0]{\@secondoftwo}%
\providecommand \translation [1]{[#1]}%
\providecommand \BibitemOpen [0]{}%
\providecommand \bibitemStop [0]{}%
\providecommand \bibitemNoStop [0]{.\EOS\space}%
\providecommand \EOS [0]{\spacefactor3000\relax}%
\providecommand \BibitemShut  [1]{\csname bibitem#1\endcsname}%
\let\auto@bib@innerbib\@empty
\bibitem [{\citenamefont {Vasala}\ and\ \citenamefont
  {Karppinen}(2015)}]{review}%
  \BibitemOpen
  \bibfield  {author} {\bibinfo {author} {\bibfnamefont {S.}~\bibnamefont
  {Vasala}}\ and\ \bibinfo {author} {\bibfnamefont {M.}~\bibnamefont
  {Karppinen}},\ }\href@noop {} {\bibfield  {journal} {\bibinfo  {journal}
  {Prog. Solid State Chem.}\ }\textbf {\bibinfo {volume} {43}},\ \bibinfo
  {pages} {1} (\bibinfo {year} {2015})}\BibitemShut {NoStop}%
\bibitem [{\citenamefont {Serrate}\ \emph {et~al.}(2007)\citenamefont
  {Serrate}, \citenamefont {{De Teresa}},\ and\ \citenamefont
  {Ibarra}}]{Serrate}%
  \BibitemOpen
  \bibfield  {author} {\bibinfo {author} {\bibfnamefont {D.}~\bibnamefont
  {Serrate}}, \bibinfo {author} {\bibfnamefont {J.~M.}\ \bibnamefont {{De
  Teresa}}}, \ and\ \bibinfo {author} {\bibfnamefont {M.~R.}\ \bibnamefont
  {Ibarra}},\ }\href@noop {} {\bibfield  {journal} {\bibinfo  {journal} {J.
  Phys. Condens. Matter.}\ }\textbf {\bibinfo {volume} {19}},\ \bibinfo {pages}
  {023201} (\bibinfo {year} {2007})}\BibitemShut {NoStop}%
\bibitem [{\citenamefont {Ferreira}\ \emph {et~al.}(2016)\citenamefont
  {Ferreira}, \citenamefont {Morrison}, \citenamefont {Yeon},\ and\
  \citenamefont {{zur Loye}}}]{LMIO}%
  \BibitemOpen
  \bibfield  {author} {\bibinfo {author} {\bibfnamefont {T.}~\bibnamefont
  {Ferreira}}, \bibinfo {author} {\bibfnamefont {G.}~\bibnamefont {Morrison}},
  \bibinfo {author} {\bibfnamefont {J.}~\bibnamefont {Yeon}}, \ and\ \bibinfo
  {author} {\bibfnamefont {H.-C.}\ \bibnamefont {{zur Loye}}},\ }\href@noop {}
  {\bibfield  {journal} {\bibinfo  {journal} {Cryst. Growth Des.}\ }\textbf
  {\bibinfo {volume} {16}},\ \bibinfo {pages} {2795} (\bibinfo {year}
  {2016})}\BibitemShut {NoStop}%
\bibitem [{\citenamefont {Kim}\ \emph {et~al.}(2015)\citenamefont {Kim},
  \citenamefont {Moon}, \citenamefont {Choi}, \citenamefont {Oh}, \citenamefont
  {Lee},\ and\ \citenamefont {Choi}}]{LCMO}%
  \BibitemOpen
  \bibfield  {author} {\bibinfo {author} {\bibfnamefont {M.~K.}\ \bibnamefont
  {Kim}}, \bibinfo {author} {\bibfnamefont {J.~Y.}\ \bibnamefont {Moon}},
  \bibinfo {author} {\bibfnamefont {H.~Y.}\ \bibnamefont {Choi}}, \bibinfo
  {author} {\bibfnamefont {S.~H.}\ \bibnamefont {Oh}}, \bibinfo {author}
  {\bibfnamefont {N.}~\bibnamefont {Lee}}, \ and\ \bibinfo {author}
  {\bibfnamefont {Y.~J.}\ \bibnamefont {Choi}},\ }\href@noop {} {\bibfield
  {journal} {\bibinfo  {journal} {J. Phys. Condens. Matter}\ }\textbf {\bibinfo
  {volume} {27}},\ \bibinfo {pages} {426002} (\bibinfo {year}
  {2015})}\BibitemShut {NoStop}%
\bibitem [{\citenamefont {S$\acute{a}$nchez-Ben$\acute{i}$tez}\ \emph
  {et~al.}(2011)\citenamefont {S$\acute{a}$nchez-Ben$\acute{i}$tez},
  \citenamefont {Mart$\acute{i}$nez-Lope}, \citenamefont {Alonso},\ and\
  \citenamefont {Garc$\acute{i}$a-Mu$\tilde{n}$oz}}]{NNMO}%
  \BibitemOpen
  \bibfield  {author} {\bibinfo {author} {\bibfnamefont {J.}~\bibnamefont
  {S$\acute{a}$nchez-Ben$\acute{i}$tez}}, \bibinfo {author} {\bibfnamefont
  {M.~J.}\ \bibnamefont {Mart$\acute{i}$nez-Lope}}, \bibinfo {author}
  {\bibfnamefont {J.~A.}\ \bibnamefont {Alonso}}, \ and\ \bibinfo {author}
  {\bibfnamefont {J.~L.}\ \bibnamefont {Garc$\acute{i}$a-Mu$\tilde{n}$oz}},\
  }\href@noop {} {\bibfield  {journal} {\bibinfo  {journal} {J. Phys. Condens.
  Matter}\ }\textbf {\bibinfo {volume} {23}},\ \bibinfo {pages} {226001}
  (\bibinfo {year} {2011})}\BibitemShut {NoStop}%
\bibitem [{\citenamefont {Kim}\ \emph {et~al.}(2009)\citenamefont {Kim},
  \citenamefont {Ohsumi}, \citenamefont {Komesu}, \citenamefont {Sakai},
  \citenamefont {Morita}, \citenamefont {Takagi},\ and\ \citenamefont
  {Arima}}]{coupling}%
  \BibitemOpen
  \bibfield  {author} {\bibinfo {author} {\bibfnamefont {B.~J.}\ \bibnamefont
  {Kim}}, \bibinfo {author} {\bibfnamefont {H.}~\bibnamefont {Ohsumi}},
  \bibinfo {author} {\bibfnamefont {T.}~\bibnamefont {Komesu}}, \bibinfo
  {author} {\bibfnamefont {S.}~\bibnamefont {Sakai}}, \bibinfo {author}
  {\bibfnamefont {T.}~\bibnamefont {Morita}}, \bibinfo {author} {\bibfnamefont
  {H.}~\bibnamefont {Takagi}}, \ and\ \bibinfo {author} {\bibfnamefont
  {T.}~\bibnamefont {Arima}},\ }\href@noop {} {\bibfield  {journal} {\bibinfo
  {journal} {Science}\ }\textbf {\bibinfo {volume} {323}},\ \bibinfo {pages}
  {1329} (\bibinfo {year} {2009})}\BibitemShut {NoStop}%
\bibitem [{\citenamefont {Kim}\ \emph {et~al.}(2008)\citenamefont {Kim},
  \citenamefont {Jin}, \citenamefont {Moon}, \citenamefont {Kim}, \citenamefont
  {Park}, \citenamefont {Leem}, \citenamefont {Yu}, \citenamefont {Noh},
  \citenamefont {Kim}, \citenamefont {Oh}, \citenamefont {Park}, \citenamefont
  {Durairaj}, \citenamefont {Cao},\ and\ \citenamefont {Rotenberg}}]{SOCL}%
  \BibitemOpen
  \bibfield  {author} {\bibinfo {author} {\bibfnamefont {B.~J.}\ \bibnamefont
  {Kim}}, \bibinfo {author} {\bibfnamefont {H.}~\bibnamefont {Jin}}, \bibinfo
  {author} {\bibfnamefont {S.~J.}\ \bibnamefont {Moon}}, \bibinfo {author}
  {\bibfnamefont {J.-Y.}\ \bibnamefont {Kim}}, \bibinfo {author} {\bibfnamefont
  {B.-G.}\ \bibnamefont {Park}}, \bibinfo {author} {\bibfnamefont {C.~S.}\
  \bibnamefont {Leem}}, \bibinfo {author} {\bibfnamefont {J.}~\bibnamefont
  {Yu}}, \bibinfo {author} {\bibfnamefont {T.~W.}\ \bibnamefont {Noh}},
  \bibinfo {author} {\bibfnamefont {C.}~\bibnamefont {Kim}}, \bibinfo {author}
  {\bibfnamefont {S.-J.}\ \bibnamefont {Oh}}, \bibinfo {author} {\bibfnamefont
  {J.-H.}\ \bibnamefont {Park}}, \bibinfo {author} {\bibfnamefont
  {V.}~\bibnamefont {Durairaj}}, \bibinfo {author} {\bibfnamefont
  {G.}~\bibnamefont {Cao}}, \ and\ \bibinfo {author} {\bibfnamefont
  {E.}~\bibnamefont {Rotenberg}},\ }\href@noop {} {\bibfield  {journal}
  {\bibinfo  {journal} {Phys. Rev. Lett.}\ }\textbf {\bibinfo {volume} {101}},\
  \bibinfo {pages} {076402} (\bibinfo {year} {2008})}\BibitemShut {NoStop}%
\bibitem [{\citenamefont {Chen}\ \emph {et~al.}(2010)\citenamefont {Chen},
  \citenamefont {Pereira},\ and\ \citenamefont {Balents}}]{SOCB}%
  \BibitemOpen
  \bibfield  {author} {\bibinfo {author} {\bibfnamefont {G.}~\bibnamefont
  {Chen}}, \bibinfo {author} {\bibfnamefont {R.}~\bibnamefont {Pereira}}, \
  and\ \bibinfo {author} {\bibfnamefont {L.}~\bibnamefont {Balents}},\
  }\href@noop {} {\bibfield  {journal} {\bibinfo  {journal} {Phys. Rev. B}\
  }\textbf {\bibinfo {volume} {82}},\ \bibinfo {pages} {174440} (\bibinfo
  {year} {2010})}\BibitemShut {NoStop}%
\bibitem [{\citenamefont {Kolchinskaya}\ \emph {et~al.}(2012)\citenamefont
  {Kolchinskaya}, \citenamefont {Komissinskiy}, \citenamefont {Yazdi},
  \citenamefont {Vafaee}, \citenamefont {Mikhailova}, \citenamefont
  {Narayanan}, \citenamefont {Ehrenberg}, \citenamefont {Wilhelm},
  \citenamefont {Rogalev},\ and\ \citenamefont {Alff}}]{XMCD12}%
  \BibitemOpen
  \bibfield  {author} {\bibinfo {author} {\bibfnamefont {A.}~\bibnamefont
  {Kolchinskaya}}, \bibinfo {author} {\bibfnamefont {P.}~\bibnamefont
  {Komissinskiy}}, \bibinfo {author} {\bibfnamefont {M.~B.}\ \bibnamefont
  {Yazdi}}, \bibinfo {author} {\bibfnamefont {M.}~\bibnamefont {Vafaee}},
  \bibinfo {author} {\bibfnamefont {D.}~\bibnamefont {Mikhailova}}, \bibinfo
  {author} {\bibfnamefont {N.}~\bibnamefont {Narayanan}}, \bibinfo {author}
  {\bibfnamefont {H.}~\bibnamefont {Ehrenberg}}, \bibinfo {author}
  {\bibfnamefont {F.}~\bibnamefont {Wilhelm}}, \bibinfo {author} {\bibfnamefont
  {A.}~\bibnamefont {Rogalev}}, \ and\ \bibinfo {author} {\bibfnamefont
  {L.}~\bibnamefont {Alff}},\ }\href@noop {} {\bibfield  {journal} {\bibinfo
  {journal} {Phys. Rev. B}\ }\textbf {\bibinfo {volume} {85}},\ \bibinfo
  {pages} {224422} (\bibinfo {year} {2012})}\BibitemShut {NoStop}%
\bibitem [{\citenamefont {Lee}\ \emph {et~al.}(2015)\citenamefont {Lee},
  \citenamefont {Sohn}, \citenamefont {Kim}, \citenamefont {Lee}, \citenamefont
  {Won}, \citenamefont {Hur}, \citenamefont {Kim}, \citenamefont {Cho},\ and\
  \citenamefont {Noh}}]{XMCD15}%
  \BibitemOpen
  \bibfield  {author} {\bibinfo {author} {\bibfnamefont {M.-C.}\ \bibnamefont
  {Lee}}, \bibinfo {author} {\bibfnamefont {C.~H.}\ \bibnamefont {Sohn}},
  \bibinfo {author} {\bibfnamefont {S.~Y.}\ \bibnamefont {Kim}}, \bibinfo
  {author} {\bibfnamefont {K.~D.}\ \bibnamefont {Lee}}, \bibinfo {author}
  {\bibfnamefont {C.~J.}\ \bibnamefont {Won}}, \bibinfo {author} {\bibfnamefont
  {N.}~\bibnamefont {Hur}}, \bibinfo {author} {\bibfnamefont {J.-Y.}\
  \bibnamefont {Kim}}, \bibinfo {author} {\bibfnamefont {D.-Y.}\ \bibnamefont
  {Cho}}, \ and\ \bibinfo {author} {\bibfnamefont {T.~W.}\ \bibnamefont
  {Noh}},\ }\href@noop {} {\bibfield  {journal} {\bibinfo  {journal} {J. Phys.
  Condens. Matter}\ }\textbf {\bibinfo {volume} {27}},\ \bibinfo {pages}
  {336002} (\bibinfo {year} {2015})}\BibitemShut {NoStop}%
\bibitem [{\citenamefont {Narayanan}\ \emph {et~al.}(2010)\citenamefont
  {Narayanan}, \citenamefont {Mikhailova}, \citenamefont {Senyshyn},
  \citenamefont {Trots}, \citenamefont {Laskowski}, \citenamefont {Blaha},
  \citenamefont {Schwarz}, \citenamefont {Fuess},\ and\ \citenamefont
  {Ehrenberg}}]{neutron}%
  \BibitemOpen
  \bibfield  {author} {\bibinfo {author} {\bibfnamefont {N.}~\bibnamefont
  {Narayanan}}, \bibinfo {author} {\bibfnamefont {D.}~\bibnamefont
  {Mikhailova}}, \bibinfo {author} {\bibfnamefont {A.}~\bibnamefont
  {Senyshyn}}, \bibinfo {author} {\bibfnamefont {D.~M.}\ \bibnamefont {Trots}},
  \bibinfo {author} {\bibfnamefont {R.}~\bibnamefont {Laskowski}}, \bibinfo
  {author} {\bibfnamefont {P.}~\bibnamefont {Blaha}}, \bibinfo {author}
  {\bibfnamefont {K.}~\bibnamefont {Schwarz}}, \bibinfo {author} {\bibfnamefont
  {H.}~\bibnamefont {Fuess}}, \ and\ \bibinfo {author} {\bibfnamefont
  {H.}~\bibnamefont {Ehrenberg}},\ }\href@noop {} {\bibfield  {journal}
  {\bibinfo  {journal} {Phys. Rev. B}\ }\textbf {\bibinfo {volume} {82}},\
  \bibinfo {pages} {024403} (\bibinfo {year} {2010})}\BibitemShut {NoStop}%
\bibitem [{\citenamefont {Song}\ \emph {et~al.}(2017)\citenamefont {Song},
  \citenamefont {Zhao}, \citenamefont {Yin}, \citenamefont {Qin}, \citenamefont
  {Zhou}, \citenamefont {Wang}, \citenamefont {Song},\ and\ \citenamefont
  {Sun}}]{Laspinglass}%
  \BibitemOpen
  \bibfield  {author} {\bibinfo {author} {\bibfnamefont {J.}~\bibnamefont
  {Song}}, \bibinfo {author} {\bibfnamefont {B.}~\bibnamefont {Zhao}}, \bibinfo
  {author} {\bibfnamefont {L.}~\bibnamefont {Yin}}, \bibinfo {author}
  {\bibfnamefont {Y.}~\bibnamefont {Qin}}, \bibinfo {author} {\bibfnamefont
  {J.}~\bibnamefont {Zhou}}, \bibinfo {author} {\bibfnamefont {D.}~\bibnamefont
  {Wang}}, \bibinfo {author} {\bibfnamefont {W.}~\bibnamefont {Song}}, \ and\
  \bibinfo {author} {\bibfnamefont {Y.}~\bibnamefont {Sun}},\ }\href@noop {}
  {\bibfield  {journal} {\bibinfo  {journal} {Dalton Trans.}\ }\textbf
  {\bibinfo {volume} {46}},\ \bibinfo {pages} {11691} (\bibinfo {year}
  {2017})}\BibitemShut {NoStop}%
\bibitem [{\citenamefont {Lutterotti}\ \emph {et~al.}(1997)\citenamefont
  {Lutterotti}, \citenamefont {Matthies}, \citenamefont {Wenk}, \citenamefont
  {Schultz},\ and\ \citenamefont {{Richardson Jr.}}}]{maud}%
  \BibitemOpen
  \bibfield  {author} {\bibinfo {author} {\bibfnamefont {L.}~\bibnamefont
  {Lutterotti}}, \bibinfo {author} {\bibfnamefont {S.}~\bibnamefont
  {Matthies}}, \bibinfo {author} {\bibfnamefont {H.-R.}\ \bibnamefont {Wenk}},
  \bibinfo {author} {\bibfnamefont {A.~S.}\ \bibnamefont {Schultz}}, \ and\
  \bibinfo {author} {\bibfnamefont {J.~W.}\ \bibnamefont {{Richardson Jr.}}},\
  }\href@noop {} {\bibfield  {journal} {\bibinfo  {journal} {J. Appl. Phys.}\
  }\textbf {\bibinfo {volume} {81}},\ \bibinfo {pages} {594} (\bibinfo {year}
  {1997})}\BibitemShut {NoStop}%
\bibitem [{\citenamefont {{Mugavero III}}\ \emph {et~al.}(2010)\citenamefont
  {{Mugavero III}}, \citenamefont {Fox}, \citenamefont {Smith},\ and\
  \citenamefont {{zur Loye}}}]{EMIO}%
  \BibitemOpen
  \bibfield  {author} {\bibinfo {author} {\bibfnamefont {S.~J.}\ \bibnamefont
  {{Mugavero III}}}, \bibinfo {author} {\bibfnamefont {A.~H.}\ \bibnamefont
  {Fox}}, \bibinfo {author} {\bibfnamefont {M.~D.}\ \bibnamefont {Smith}}, \
  and\ \bibinfo {author} {\bibfnamefont {H.-C.}\ \bibnamefont {{zur Loye}}},\
  }\href@noop {} {\bibfield  {journal} {\bibinfo  {journal} {J. Solid State
  Chem.}\ }\textbf {\bibinfo {volume} {183}},\ \bibinfo {pages} {465} (\bibinfo
  {year} {2010})}\BibitemShut {NoStop}%
\bibitem [{\citenamefont {Zhang}\ \emph {et~al.}(1993)\citenamefont {Zhang},
  \citenamefont {{von Dreele}},\ and\ \citenamefont {Eyring}}]{Tb11O20}%
  \BibitemOpen
  \bibfield  {author} {\bibinfo {author} {\bibfnamefont {J.}~\bibnamefont
  {Zhang}}, \bibinfo {author} {\bibfnamefont {R.~B.}\ \bibnamefont {{von
  Dreele}}}, \ and\ \bibinfo {author} {\bibfnamefont {L.}~\bibnamefont
  {Eyring}},\ }\href@noop {} {\bibfield  {journal} {\bibinfo  {journal} {J.
  Solid State Chem.}\ }\textbf {\bibinfo {volume} {104}},\ \bibinfo {pages}
  {21} (\bibinfo {year} {1993})}\BibitemShut {NoStop}%
\bibitem [{\citenamefont {Baran}\ \emph {et~al.}(2013)\citenamefont {Baran},
  \citenamefont {Duraj}, \citenamefont {Hoser}, \citenamefont {Penc},\ and\
  \citenamefont {Szytu{\l}a}}]{Tb11O20m}%
  \BibitemOpen
  \bibfield  {author} {\bibinfo {author} {\bibfnamefont {S.}~\bibnamefont
  {Baran}}, \bibinfo {author} {\bibfnamefont {R.}~\bibnamefont {Duraj}},
  \bibinfo {author} {\bibfnamefont {A.}~\bibnamefont {Hoser}}, \bibinfo
  {author} {\bibfnamefont {B.}~\bibnamefont {Penc}}, \ and\ \bibinfo {author}
  {\bibfnamefont {A.}~\bibnamefont {Szytu{\l}a}},\ }\href@noop {} {\bibfield
  {journal} {\bibinfo  {journal} {Acta Phys. Pol. A}\ }\textbf {\bibinfo
  {volume} {123}},\ \bibinfo {pages} {98} (\bibinfo {year} {2013})}\BibitemShut
  {NoStop}%
\bibitem [{\citenamefont {Koehler}\ \emph {et~al.}(1958)\citenamefont
  {Koehler}, \citenamefont {Wollan},\ and\ \citenamefont {Wilkinson}}]{Ho2O3}%
  \BibitemOpen
  \bibfield  {author} {\bibinfo {author} {\bibfnamefont {W.~C.}\ \bibnamefont
  {Koehler}}, \bibinfo {author} {\bibfnamefont {E.~O.}\ \bibnamefont {Wollan}},
  \ and\ \bibinfo {author} {\bibfnamefont {M.~K.}\ \bibnamefont {Wilkinson}},\
  }\href@noop {} {\bibfield  {journal} {\bibinfo  {journal} {Phys. Rev.}\
  }\textbf {\bibinfo {volume} {110}},\ \bibinfo {pages} {37} (\bibinfo {year}
  {1958})}\BibitemShut {NoStop}%
\bibitem [{\citenamefont {Boutahar}\ \emph {et~al.}(2017)\citenamefont
  {Boutahar}, \citenamefont {Moubah}, \citenamefont {Hlil}, \citenamefont
  {Lassri},\ and\ \citenamefont {Lorenzo}}]{Ho2O3m}%
  \BibitemOpen
  \bibfield  {author} {\bibinfo {author} {\bibfnamefont {A.}~\bibnamefont
  {Boutahar}}, \bibinfo {author} {\bibfnamefont {R.}~\bibnamefont {Moubah}},
  \bibinfo {author} {\bibfnamefont {E.~K.}\ \bibnamefont {Hlil}}, \bibinfo
  {author} {\bibfnamefont {H.}~\bibnamefont {Lassri}}, \ and\ \bibinfo {author}
  {\bibfnamefont {E.}~\bibnamefont {Lorenzo}},\ }\href@noop {} {\bibfield
  {journal} {\bibinfo  {journal} {Sci. Rep.}\ }\textbf {\bibinfo {volume}
  {7}},\ \bibinfo {pages} {13904} (\bibinfo {year} {2017})}\BibitemShut
  {NoStop}%
\bibitem [{\citenamefont {{Van Vleck}}(1932)}]{Vleck}%
  \BibitemOpen
  \bibfield  {author} {\bibinfo {author} {\bibfnamefont {J.~H.}\ \bibnamefont
  {{Van Vleck}}},\ }\href@noop {} {\emph {\bibinfo {title} {The Theory of
  Electric and Magnetic Susceptibilities}}}\ (\bibinfo  {publisher} {Oxford
  University Press},\ \bibinfo {year} {1932})\ p.\ \bibinfo {pages}
  {226}\BibitemShut {NoStop}%
\bibitem [{\citenamefont {Takikawa}\ \emph {et~al.}(2010)\citenamefont
  {Takikawa}, \citenamefont {Ebisu},\ and\ \citenamefont {Nagata}}]{EuBO3}%
  \BibitemOpen
  \bibfield  {author} {\bibinfo {author} {\bibfnamefont {Y.}~\bibnamefont
  {Takikawa}}, \bibinfo {author} {\bibfnamefont {S.}~\bibnamefont {Ebisu}}, \
  and\ \bibinfo {author} {\bibfnamefont {S.}~\bibnamefont {Nagata}},\
  }\href@noop {} {\bibfield  {journal} {\bibinfo  {journal} {J. Phys. Chem.
  Solids}\ }\textbf {\bibinfo {volume} {71}},\ \bibinfo {pages} {1592}
  (\bibinfo {year} {2010})}\BibitemShut {NoStop}%
\bibitem [{\citenamefont {Samata}\ \emph {et~al.}(2015)\citenamefont {Samata},
  \citenamefont {Wada},\ and\ \citenamefont {Ozawa}}]{EuOOH}%
  \BibitemOpen
  \bibfield  {author} {\bibinfo {author} {\bibfnamefont {H.}~\bibnamefont
  {Samata}}, \bibinfo {author} {\bibfnamefont {N.}~\bibnamefont {Wada}}, \ and\
  \bibinfo {author} {\bibfnamefont {T.~C.}\ \bibnamefont {Ozawa}},\ }\href@noop
  {} {\bibfield  {journal} {\bibinfo  {journal} {J. Rare Earths}\ }\textbf
  {\bibinfo {volume} {33}},\ \bibinfo {pages} {177} (\bibinfo {year}
  {2015})}\BibitemShut {NoStop}%
\bibitem [{\citenamefont {Y$\acute{a}\tilde{n}$ez-Vilar}\ \emph
  {et~al.}(2011)\citenamefont {Y$\acute{a}\tilde{n}$ez-Vilar}, \citenamefont
  {Mun}, \citenamefont {Zapf}, \citenamefont {Ueland}, \citenamefont {Gardner},
  \citenamefont {Thompson}, \citenamefont {Singleton}, \citenamefont
  {S$\acute{a}$nchez-And$\acute{u}$jar}, \citenamefont {Mira}, \citenamefont
  {Biskup}, \citenamefont {Se$\tilde{n}$ar$\acute{i}$s-Rodr$\acute{i}$guez},\
  and\ \citenamefont {Batista}}]{LMCO}%
  \BibitemOpen
  \bibfield  {author} {\bibinfo {author} {\bibfnamefont {S.}~\bibnamefont
  {Y$\acute{a}\tilde{n}$ez-Vilar}}, \bibinfo {author} {\bibfnamefont {E.~D.}\
  \bibnamefont {Mun}}, \bibinfo {author} {\bibfnamefont {V.~S.}\ \bibnamefont
  {Zapf}}, \bibinfo {author} {\bibfnamefont {B.~G.}\ \bibnamefont {Ueland}},
  \bibinfo {author} {\bibfnamefont {J.~S.}\ \bibnamefont {Gardner}}, \bibinfo
  {author} {\bibfnamefont {J.~D.}\ \bibnamefont {Thompson}}, \bibinfo {author}
  {\bibfnamefont {J.}~\bibnamefont {Singleton}}, \bibinfo {author}
  {\bibfnamefont {M.}~\bibnamefont {S$\acute{a}$nchez-And$\acute{u}$jar}},
  \bibinfo {author} {\bibfnamefont {J.}~\bibnamefont {Mira}}, \bibinfo {author}
  {\bibfnamefont {N.}~\bibnamefont {Biskup}}, \bibinfo {author} {\bibfnamefont
  {M.~A.}\ \bibnamefont {Se$\tilde{n}$ar$\acute{i}$s-Rodr$\acute{i}$guez}}, \
  and\ \bibinfo {author} {\bibfnamefont {C.~D.}\ \bibnamefont {Batista}},\
  }\href@noop {} {\bibfield  {journal} {\bibinfo  {journal} {Phys. Rev. B}\
  }\textbf {\bibinfo {volume} {84}},\ \bibinfo {pages} {134427} (\bibinfo
  {year} {2011})}\BibitemShut {NoStop}%
\bibitem [{\citenamefont {Blasco}\ \emph {et~al.}(2016)\citenamefont {Blasco},
  \citenamefont {Garc$\acute{i}$a}, \citenamefont {Sub$\acute{i}$as},
  \citenamefont {Stankiewicz}, \citenamefont
  {Rodr$\acute{i}$guez-Velamaz$\acute{a}$n}, \citenamefont {Ritter},
  \citenamefont {Garc$\acute{i}$a-Mu$\tilde{n}$oz},\ and\ \citenamefont
  {Fauth}}]{YCMO}%
  \BibitemOpen
  \bibfield  {author} {\bibinfo {author} {\bibfnamefont {J.}~\bibnamefont
  {Blasco}}, \bibinfo {author} {\bibfnamefont {J.}~\bibnamefont
  {Garc$\acute{i}$a}}, \bibinfo {author} {\bibfnamefont {G.}~\bibnamefont
  {Sub$\acute{i}$as}}, \bibinfo {author} {\bibfnamefont {J.}~\bibnamefont
  {Stankiewicz}}, \bibinfo {author} {\bibfnamefont {J.~A.}\ \bibnamefont
  {Rodr$\acute{i}$guez-Velamaz$\acute{a}$n}}, \bibinfo {author} {\bibfnamefont
  {C.}~\bibnamefont {Ritter}}, \bibinfo {author} {\bibfnamefont {J.~L.}\
  \bibnamefont {Garc$\acute{i}$a-Mu$\tilde{n}$oz}}, \ and\ \bibinfo {author}
  {\bibfnamefont {F.}~\bibnamefont {Fauth}},\ }\href@noop {} {\bibfield
  {journal} {\bibinfo  {journal} {Phys. Rev. B}\ }\textbf {\bibinfo {volume}
  {93}},\ \bibinfo {pages} {214401} (\bibinfo {year} {2016})}\BibitemShut
  {NoStop}%
\bibitem [{\citenamefont {Perry}\ \emph {et~al.}(2001)\citenamefont {Perry},
  \citenamefont {Galvin}, \citenamefont {Grigera}, \citenamefont {Capogna},
  \citenamefont {J.Schofield}, \citenamefont {Mackenzie}, \citenamefont
  {Chiao}, \citenamefont {Julian}, \citenamefont {Ikeda}, \citenamefont
  {Nakatsuji}, \citenamefont {Maeno},\ and\ \citenamefont {Pfleiderer}}]{flop}%
  \BibitemOpen
  \bibfield  {author} {\bibinfo {author} {\bibfnamefont {R.~S.}\ \bibnamefont
  {Perry}}, \bibinfo {author} {\bibfnamefont {L.~M.}\ \bibnamefont {Galvin}},
  \bibinfo {author} {\bibfnamefont {S.~A.}\ \bibnamefont {Grigera}}, \bibinfo
  {author} {\bibfnamefont {L.}~\bibnamefont {Capogna}}, \bibinfo {author}
  {\bibfnamefont {A.}~\bibnamefont {J.Schofield}}, \bibinfo {author}
  {\bibfnamefont {A.~P.}\ \bibnamefont {Mackenzie}}, \bibinfo {author}
  {\bibfnamefont {M.}~\bibnamefont {Chiao}}, \bibinfo {author} {\bibfnamefont
  {S.~R.}\ \bibnamefont {Julian}}, \bibinfo {author} {\bibfnamefont {S.~I.}\
  \bibnamefont {Ikeda}}, \bibinfo {author} {\bibfnamefont {S.}~\bibnamefont
  {Nakatsuji}}, \bibinfo {author} {\bibfnamefont {Y.}~\bibnamefont {Maeno}}, \
  and\ \bibinfo {author} {\bibfnamefont {C.}~\bibnamefont {Pfleiderer}},\
  }\href@noop {} {\bibfield  {journal} {\bibinfo  {journal} {Phys. Rev. Lett.}\
  }\textbf {\bibinfo {volume} {86}},\ \bibinfo {pages} {2661} (\bibinfo {year}
  {2001})}\BibitemShut {NoStop}%
\bibitem [{\citenamefont {Horner}\ and\ \citenamefont
  {Varma}(1968)}]{reorientation}%
  \BibitemOpen
  \bibfield  {author} {\bibinfo {author} {\bibfnamefont {H.}~\bibnamefont
  {Horner}}\ and\ \bibinfo {author} {\bibfnamefont {C.~M.}\ \bibnamefont
  {Varma}},\ }\href@noop {} {\bibfield  {journal} {\bibinfo  {journal} {Phys.
  Rev. Lett.}\ }\textbf {\bibinfo {volume} {20}},\ \bibinfo {pages} {16}
  (\bibinfo {year} {1968})}\BibitemShut {NoStop}%
\bibitem [{\citenamefont {Zhong}\ \emph {et~al.}(2013)\citenamefont {Zhong},
  \citenamefont {Au},\ and\ \citenamefont {Du}}]{CPB}%
  \BibitemOpen
  \bibfield  {author} {\bibinfo {author} {\bibfnamefont {W.}~\bibnamefont
  {Zhong}}, \bibinfo {author} {\bibfnamefont {C.-T.}\ \bibnamefont {Au}}, \
  and\ \bibinfo {author} {\bibfnamefont {Y.-W.}\ \bibnamefont {Du}},\
  }\href@noop {} {\bibfield  {journal} {\bibinfo  {journal} {Chin. Phys. B}\
  }\textbf {\bibinfo {volume} {22}},\ \bibinfo {pages} {057501} (\bibinfo
  {year} {2013})}\BibitemShut {NoStop}%
\bibitem [{\citenamefont {Currie}\ \emph {et~al.}(1995)\citenamefont {Currie},
  \citenamefont {Vente}, \citenamefont {Frikkee},\ and\ \citenamefont
  {IJdo}}]{LNIO}%
  \BibitemOpen
  \bibfield  {author} {\bibinfo {author} {\bibfnamefont {R.~C.}\ \bibnamefont
  {Currie}}, \bibinfo {author} {\bibfnamefont {J.~F.}\ \bibnamefont {Vente}},
  \bibinfo {author} {\bibfnamefont {E.}~\bibnamefont {Frikkee}}, \ and\
  \bibinfo {author} {\bibfnamefont {D.~J.~W.}\ \bibnamefont {IJdo}},\
  }\href@noop {} {\bibfield  {journal} {\bibinfo  {journal} {J. Solid State
  Chem.}\ }\textbf {\bibinfo {volume} {116}},\ \bibinfo {pages} {199} (\bibinfo
  {year} {1995})}\BibitemShut {NoStop}%
\bibitem [{\citenamefont {Bufai{\c c}al}\ \emph {et~al.}(2014)\citenamefont
  {Bufai{\c c}al}, \citenamefont {Adriano}, \citenamefont {Lora-Serrano},
  \citenamefont {Duque}, \citenamefont {Mendon{\c c}a-Ferreira}, \citenamefont
  {Rojas-Ayala}, \citenamefont {Baggio-Saitovitch}, \citenamefont {Bittar},\
  and\ \citenamefont {Pagliuso}}]{LFIO2}%
  \BibitemOpen
  \bibfield  {author} {\bibinfo {author} {\bibfnamefont {L.}~\bibnamefont
  {Bufai{\c c}al}}, \bibinfo {author} {\bibfnamefont {C.}~\bibnamefont
  {Adriano}}, \bibinfo {author} {\bibfnamefont {R.}~\bibnamefont
  {Lora-Serrano}}, \bibinfo {author} {\bibfnamefont {J.~G.~S.}\ \bibnamefont
  {Duque}}, \bibinfo {author} {\bibfnamefont {L.}~\bibnamefont {Mendon{\c
  c}a-Ferreira}}, \bibinfo {author} {\bibfnamefont {C.}~\bibnamefont
  {Rojas-Ayala}}, \bibinfo {author} {\bibfnamefont {E.}~\bibnamefont
  {Baggio-Saitovitch}}, \bibinfo {author} {\bibfnamefont {E.~M.}\ \bibnamefont
  {Bittar}}, \ and\ \bibinfo {author} {\bibfnamefont {P.~G.}\ \bibnamefont
  {Pagliuso}},\ }\href@noop {} {\bibfield  {journal} {\bibinfo  {journal} {J.
  Solid State Chem.}\ }\textbf {\bibinfo {volume} {212}},\ \bibinfo {pages}
  {23} (\bibinfo {year} {2014})}\BibitemShut {NoStop}%
\bibitem [{\citenamefont {Bufai{\c c}al}\ \emph {et~al.}(2008)\citenamefont
  {Bufai{\c c}al}, \citenamefont {Mendon{\c c}a-Ferreira}, \citenamefont
  {Lora-Serrano}, \citenamefont {Ag{\"u}ero}, \citenamefont {Torriani},
  \citenamefont {Granado}, \citenamefont {Pagliuso}, \citenamefont {Caytuero},\
  and\ \citenamefont {Baggio-Saitovich}}]{LFIO}%
  \BibitemOpen
  \bibfield  {author} {\bibinfo {author} {\bibfnamefont {L.}~\bibnamefont
  {Bufai{\c c}al}}, \bibinfo {author} {\bibfnamefont {L.}~\bibnamefont
  {Mendon{\c c}a-Ferreira}}, \bibinfo {author} {\bibfnamefont {R.}~\bibnamefont
  {Lora-Serrano}}, \bibinfo {author} {\bibfnamefont {O.}~\bibnamefont
  {Ag{\"u}ero}}, \bibinfo {author} {\bibfnamefont {I.}~\bibnamefont
  {Torriani}}, \bibinfo {author} {\bibfnamefont {E.}~\bibnamefont {Granado}},
  \bibinfo {author} {\bibfnamefont {P.~G.}\ \bibnamefont {Pagliuso}}, \bibinfo
  {author} {\bibfnamefont {A.}~\bibnamefont {Caytuero}}, \ and\ \bibinfo
  {author} {\bibfnamefont {E.}~\bibnamefont {Baggio-Saitovich}},\ }\href@noop
  {} {\bibfield  {journal} {\bibinfo  {journal} {J. Appl. Phys.}\ }\textbf
  {\bibinfo {volume} {103}},\ \bibinfo {pages} {07F716} (\bibinfo {year}
  {2008})}\BibitemShut {NoStop}%
\bibitem [{\citenamefont {Choy}\ \emph {et~al.}(1995)\citenamefont {Choy},
  \citenamefont {Kim}, \citenamefont {Hwang}, \citenamefont {Demazeau},\ and\
  \citenamefont {Jung}}]{LLIO}%
  \BibitemOpen
  \bibfield  {author} {\bibinfo {author} {\bibfnamefont {J.-H.}\ \bibnamefont
  {Choy}}, \bibinfo {author} {\bibfnamefont {D.-K.}\ \bibnamefont {Kim}},
  \bibinfo {author} {\bibfnamefont {S.-H.}\ \bibnamefont {Hwang}}, \bibinfo
  {author} {\bibfnamefont {G.}~\bibnamefont {Demazeau}}, \ and\ \bibinfo
  {author} {\bibfnamefont {D.-Y.}\ \bibnamefont {Jung}},\ }\href@noop {}
  {\bibfield  {journal} {\bibinfo  {journal} {J. Am. Chem. Soc.}\ }\textbf
  {\bibinfo {volume} {117}},\ \bibinfo {pages} {8557} (\bibinfo {year}
  {1995})}\BibitemShut {NoStop}%
\bibitem [{\citenamefont {Wakeshima}\ \emph {et~al.}(1999)\citenamefont
  {Wakeshima}, \citenamefont {Harada},\ and\ \citenamefont {Hinatsu}}]{SYIO99}%
  \BibitemOpen
  \bibfield  {author} {\bibinfo {author} {\bibfnamefont {M.}~\bibnamefont
  {Wakeshima}}, \bibinfo {author} {\bibfnamefont {D.}~\bibnamefont {Harada}}, \
  and\ \bibinfo {author} {\bibfnamefont {Y.~J.}\ \bibnamefont {Hinatsu}},\
  }\href@noop {} {\bibfield  {journal} {\bibinfo  {journal} {J. Alloys Compd.}\
  }\textbf {\bibinfo {volume} {287}},\ \bibinfo {pages} {130} (\bibinfo {year}
  {1999})}\BibitemShut {NoStop}%
\bibitem [{\citenamefont {Corredor}\ \emph {et~al.}(2017)\citenamefont
  {Corredor}, \citenamefont {Aslan-Cansever}, \citenamefont {Sturza},
  \citenamefont {Manna}, \citenamefont {Maljuk}, \citenamefont {Gass},
  \citenamefont {Dey}, \citenamefont {Wolter}, \citenamefont {Kataeva},
  \citenamefont {Zimmermann}, \citenamefont {Geyer}, \citenamefont {Blum},
  \citenamefont {Wurmehl},\ and\ \citenamefont {B$\ddot{u}$chner}}]{SYIO17}%
  \BibitemOpen
  \bibfield  {author} {\bibinfo {author} {\bibfnamefont {L.~T.}\ \bibnamefont
  {Corredor}}, \bibinfo {author} {\bibfnamefont {G.}~\bibnamefont
  {Aslan-Cansever}}, \bibinfo {author} {\bibfnamefont {M.}~\bibnamefont
  {Sturza}}, \bibinfo {author} {\bibfnamefont {K.}~\bibnamefont {Manna}},
  \bibinfo {author} {\bibfnamefont {A.}~\bibnamefont {Maljuk}}, \bibinfo
  {author} {\bibfnamefont {S.}~\bibnamefont {Gass}}, \bibinfo {author}
  {\bibfnamefont {T.}~\bibnamefont {Dey}}, \bibinfo {author} {\bibfnamefont
  {A.~U.~B.}\ \bibnamefont {Wolter}}, \bibinfo {author} {\bibfnamefont
  {O.}~\bibnamefont {Kataeva}}, \bibinfo {author} {\bibfnamefont
  {A.}~\bibnamefont {Zimmermann}}, \bibinfo {author} {\bibfnamefont
  {M.}~\bibnamefont {Geyer}}, \bibinfo {author} {\bibfnamefont {C.~G.~F.}\
  \bibnamefont {Blum}}, \bibinfo {author} {\bibfnamefont {S.}~\bibnamefont
  {Wurmehl}}, \ and\ \bibinfo {author} {\bibfnamefont {B.}~\bibnamefont
  {B$\ddot{u}$chner}},\ }\href@noop {} {\bibfield  {journal} {\bibinfo
  {journal} {Phys. Rev. B}\ }\textbf {\bibinfo {volume} {95}},\ \bibinfo
  {pages} {064418} (\bibinfo {year} {2017})}\BibitemShut {NoStop}%
\bibitem [{\citenamefont {Mikhailova}\ \emph {et~al.}(2010)\citenamefont
  {Mikhailova}, \citenamefont {Narayanan}, \citenamefont {Gruner},
  \citenamefont {Voss}, \citenamefont {Senyshyn}, \citenamefont {Trots},
  \citenamefont {Fuess},\ and\ \citenamefont {Ehrenberg}}]{S3CIO}%
  \BibitemOpen
  \bibfield  {author} {\bibinfo {author} {\bibfnamefont {D.}~\bibnamefont
  {Mikhailova}}, \bibinfo {author} {\bibfnamefont {N.}~\bibnamefont
  {Narayanan}}, \bibinfo {author} {\bibfnamefont {W.}~\bibnamefont {Gruner}},
  \bibinfo {author} {\bibfnamefont {A.}~\bibnamefont {Voss}}, \bibinfo {author}
  {\bibfnamefont {A.}~\bibnamefont {Senyshyn}}, \bibinfo {author}
  {\bibfnamefont {D.~M.}\ \bibnamefont {Trots}}, \bibinfo {author}
  {\bibfnamefont {H.}~\bibnamefont {Fuess}}, \ and\ \bibinfo {author}
  {\bibfnamefont {H.}~\bibnamefont {Ehrenberg}},\ }\href@noop {} {\bibfield
  {journal} {\bibinfo  {journal} {Inorg. Chem.}\ }\textbf {\bibinfo {volume}
  {49}},\ \bibinfo {pages} {10348} (\bibinfo {year} {2010})}\BibitemShut
  {NoStop}%
\bibitem [{\citenamefont {Nguyen}\ and\ \citenamefont {{zur
  Loye}}(1995)}]{chains}%
  \BibitemOpen
  \bibfield  {author} {\bibinfo {author} {\bibfnamefont {T.~N.}\ \bibnamefont
  {Nguyen}}\ and\ \bibinfo {author} {\bibfnamefont {H.-C.}\ \bibnamefont {{zur
  Loye}}},\ }\href@noop {} {\bibfield  {journal} {\bibinfo  {journal} {J. Solid
  State Chem.}\ }\textbf {\bibinfo {volume} {117}},\ \bibinfo {pages} {300}
  (\bibinfo {year} {1995})}\BibitemShut {NoStop}%
\end{thebibliography}%

\end{document}